\documentclass[osajnl,twocolumn,showpacs, superscriptaddress,10pt]{revtex4-1}

\usepackage{amsmath,amssymb,graphicx}
\usepackage{url}
\usepackage{braket}

\begin{document}

\title{Polarization-entangled photon-pair generation in commercial-grade polarization-maintaining fiber}

\author{Bin Fang}\email{Corresponding author: bfang@udel.edu}
\affiliation{Department of Physics and Astronomy, University of Delaware, Newark, DE 19716, USA}

\author{Offir Cohen}
\affiliation{Department of Physics and Astronomy, University of Delaware, Newark, DE 19716, USA}
\affiliation{Present address: Joint Quantum Institute, National Institute of Standards and Technology \& \\University of Maryland, Gaithersburg, MD 20849, USA}

\author{Virginia O. Lorenz}
\affiliation{Department of Physics and Astronomy, University of Delaware, Newark, DE 19716, USA}

\begin{abstract}
We demonstrate a fiber-based source of polarization-entangled photon pairs at visible wavelengths suitable for integration with local quantum processing schemes. The photons are created through birefringent phase-matching in spontaneous four-wave mixing inside a Sagnac interferometer. We address entanglement degradation due to temporal distinguishability of the photons to enable the generation of a spectrally unfiltered polarization-entangled photon-pair state with $95.86\pm0.10\%$ fidelity to a maximally entangled Bell state, evaluated with a tomographic state reconstruction without applying any corrections or background subtractions. Owing to the large birefringence of the fiber, photons are created far detuned from the pump, where Raman contamination is negligible. This source's spatial mode and ability to produce spectrally uncorrelated photons make it suitable for implementing quantum information protocols over free-space and fiber-based networks.
\end{abstract}

\maketitle

\section{Introduction}
Entanglement between photons is a useful resource for many applications such as quantum communication protocols \cite{PhysRevLett.67.661}, quantum computation schemes \cite{knill2001scheme} and fundamental tests of quantum mechanics \cite{weihs2001photonic}. Powerful techniques to create polarization-entangled photons have been demonstrated using the nonlinear process of spontaneous parametric down-conversion (SPDC) in bulk crystals \cite{PhysRevLett.75.4337, PhysRevA.74.013802}, periodically polled crystals \cite{kuzucu2008pulsed, sauge2007narrowband, Hubel:OptExpress:2007} and waveguides \cite{takesue2005generation,lim2008stable,zhong2010high,Martin:NewJournalOfPhysics:2010,Kaiser:NewJournalOfPhysics:2012}. However, spatial entanglement and/or the non-Gaussian spatial mode of the photons generated in these media normally reduce the efficiency of interfacing with free-space quantum operations such as storage in atomic vapor quantum memories and of coupling into single-mode silica fiber for distribution. Thus, these polarization-entangled photon-pair sources are not inherently matched for direct integration into quantum fiber- or free-space-based protocols. Photons generated in optical fiber, on the other hand, are in a single spatial mode that is suitable for coupling into networks composed of similar fibers. With this spirit in mind, generation of polarization-entangled photons in photonic crystal fibers (PCFs) via spontaneous four-wave mixing (SFWM) was demonstrated \cite{ PhysRevA.76.043836,PhysRevLett.99.120501}. Nonetheless, the spatial mode of the photons emerging from PCFs is not perfectly Gaussian due to the complex hexagonal shape of these fibers, and hence is not perfectly compatible with single-mode fibers (SMFs) without special treatment \cite{ling2009mode}. Generation of entangled photons at telecom wavelengths has been demonstrated in conventional fibers such as dispersion-shifted fibers (DSF) \cite{PhysRevA.70.031802,Li:PhysicalReviewLetters:2005,Medic:OptLett:2010,zhoupolarization,Zhou:Arxiv13077207QuantPh:2013} and SMFs \cite{hall2009drop,Hall:PhysicalReviewLetters:2011}, thus enabling the integration of the source with a distribution network composed of fibers. One of the difficulties arising from photon-pair generation in optical fibers, including DSF and SMF, is the Raman scattering that occurs simultaneously with the SFWM process and adds background photons (mainly at the idler wavelength), and thus harms the pair-wise emission nature of the source. Cooling the fiber to 77 K \cite{Medic:OptLett:2010} has been shown to effectively reduce Raman background, but doing so adds to setup complications. 

Recently, it was demonstrated that efficient generation of photon-pairs is possible using standard, commercially-available polarization-maintaining fibers (PMFs) \cite{Smith:OptExpress:2009} by employing the so-called birefringent phase-matching, in which the photons are produced with polarization orthogonal to the pump beam. The large birefringence ($\sim3.5\times10^{-4}$) of PMF yields a $\sim60\, \text{THz}$ detuning of the photons' phase-matched wavelengths from the pump, thus almost eliminating Raman contamination \cite{stolen1981phase}. It has been predicted\,\cite{Smith:OptExpress:2009,Fang:OptExpress:2013} and demonstrated\,\cite{PhysRevA.83.031806} that such a source can generate the photons with no spectral correlations between them. Ref.\,\cite{PhysRevA.83.031806} also demonstrated that the pure-wavepacket heralded photons could be coupled into single-mode fibers with 85\% efficiency, showing the potential of the source for quantum information processing applications. In addition, the photon pairs are created at visible wavelengths and are tunable over a broad spectrum, meaning that: 1) Commercially available and cheap silicon-based single photon detectors can be used. 2) The photons can be generated at wavelengths suitable for storage in already demonstrated quantum memories \cite{Simon2010,Hammerer2010}, which are intrinsically compatible only with specific (usually visible) wavelengths due to the utilized atomic transitions. For these reasons, it is appealing to use a PMF to generate polarization-entangled photons: such a source can be incorporated into free-space or fiber networks that distribute the photons for use in various protocols, including protocols that require photon storage and retrieval, while cheap and efficient detectors can be used for projective measurements and/or state analysis. A recently demonstrated spliced source \cite{meyer2012generating} has the advantage of single-path entangled photon generation, but relies on the identicality of the two spliced pieces, whereas experimental observations \cite{PhysRevA.83.031806} show that such identicality is not necessarily the case. In this paper, we demonstrate and investigate a polarization-entangled photon-pair source based on PMF in a Sagnac loop \cite{PhysRevA.69.041801}. We experimentally study entanglement degradation due to imperfect temporal balance in the interferometer, highlighting the need for careful attention to temporal mode-matching when dealing with femtosecond sources. We show that when imbalance is compensated, highly entangled photon-pairs are created without filtering the photons spectrally. The ability to generate entangled photons without loss, including loss imposed by filters, is crucial for scaling up the implementation of various quantum-based protocols and for demonstrations of loophole-free tests of Bell's inequality.

\section{Background}
In the SFWM process, two pump photons at angular frequency $\omega_{p}$ are annihilated by a $\chi^{(3)}$ medium to generate two side-band photons at angular frequencies $\omega_{s}$ and $\omega_{i}$, denoted as signal and idler, respectively ($\omega_{s} > \omega_{i}$). Energy conservation and phase-matching have to be satisfied; for a standard polarization-maintaining fiber these conditions can be written as \cite{Smith:OptExpress:2009}
\begin{subequations}\begin{gather}
2\omega_{p}=\omega_{s}+\omega_{i},\label{eq:match1}
\\
\Delta k=2k(\omega_{p})-k(\omega_{s})-k(\omega_{i})+2\Delta n\frac{\omega_{p}}{c}=0,\label{eq:match2}
\end{gather}\end{subequations} 
where $\omega_{p,s,i}$ are the angular frequencies of the pump, signal and idler, respectively, $k(\omega)$ is the wave vector of the angular frequency $\omega$ given by the dispersion of pure silica glass, $\Delta k$ is the phase mismatch and $\Delta n$ is the birefringence of the PMF. For simplicity, we neglect here the effect of self- and cross-phase modulation, which can be justified when the pump power is low enough. Note that the fulfillment of the phase-matching condition Eqs.~(1) relies on the birefringence of the fiber and can occur for pump wavelengths far detuned from the zero-dispersion wavelength\,\cite{Smith:OptExpress:2009} (as opposed to the traditional way of generating photon pairs in optical fibers in which phase-matching is achieved by using a pump at wavelengths in the vicinity of the fiber's zero dispersion wavelength \cite{ Li:PhysicalReviewLetters:2005,Medic:OptLett:2010,hall2009drop,Hall:PhysicalReviewLetters:2011,PhysRevA.76.043836,PhysRevLett.99.120501, harvey2003scalar}). As shown in Fig.~\ref{exp}(a), in birefringent phase-matching the pump propagates on the slow axis of the fiber ($\Delta n>0$),  and signal/idler photons are created with polarization orthogonal to the pump, traveling on the fast axis of the fiber. In order to generate entanglement, the PMF is inserted into a Sagnac interferometer (see the schematic of the Sagnac loop in Fig.~\ref{exp}(b)): The pump, whose polarization is oriented at $45^{\circ}$, is split equally by a polarizing beam splitter (PBS1). The horizontal component is transmitted and then coupled into one end of the PMF to generate photon pairs with vertical polarization. The other end of the fiber is twisted by $90^{\circ}$ such that after exiting the fiber the photon pairs are transmitted by PBS1 with horizontal polarization. This photon-pair state can be written as $\Ket{\psi}_{HH}=\Ket{\phi_{HH}}\otimes\Ket{H_{s},H_{i}}$, where $\Ket{\phi_{\alpha \beta}}$ ($\alpha,\beta = H, V$) denotes the photon-pair spectral and spatial degrees of freedom, with $H$ representing horizontal and $V$ vertical polarization, and $\Ket{\alpha_{s},\beta_{i}}$ is the polarization state for the signal ($\alpha_s$) and idler ($\beta_i$). Similarly, the vertical component of the pump is coupled into the other end of the PMF (note that because the PMF is twisted by $90^{\circ}$, both pump components are launched on the slow axis of the PMF and produce photons on the fast axis) and can create a pair with vertical polarization that is reflected by PBS1. Both output states are combined together by PBS1. Upon the creation of a photon pair (and the interaction's lowest order produces only one pair), it is impossible to tell whether they were generated in the clockwise or counter-clockwise path; thus, the two paths interfere, resulting in the superposition of the two states:
\begin{equation}
\Ket{\psi}=\frac{1}{\sqrt{2}}(\Ket{\phi_{HH}}\otimes\Ket{H_{s}H_{i}}+\exp{(i\varphi)}\Ket{\phi_{VV}}\otimes\Ket{V_{s}V_{i}})
\end{equation}
where $\varphi$ is the relative phase between the two states. Ideally, the Sagnac loop is perfectly symmetric, and the outputs of the two counter-propagating paths are spatially and temporally indistinguishable such that $\Ket{\phi_{HH}}=\Ket{\phi_{VV}}$; thus, the polarization-entangled state   $\Ket{\Psi}=(\Ket{H_{s}H_{i}}+\exp{(i\varphi)}\Ket{V_{s}V_{i}})/\sqrt{2}$ is created. In reality, however, we always expect some imperfections to exist that result in distinguishing information between the paths. Consequently, the polarization state is not described by a pure wave-function but by the density matrix,
\begin{align}
\rho = & \frac{1}{2}(\Ket{H_{s}H_{i}}\Bra{H_{s}H_{i}}+\Ket{V_{s}V_{i}}\Bra{V_{s}V_{i}}\nonumber \\
 & +\exp{(i\varphi)}\Braket{\phi_{VV}|\phi_{HH}}\Ket{H_{s}H_{i}}\Bra{V_{s}V_{i}} \\
 & + \exp{(-i\varphi)}\Braket{\phi_{HH}|\phi_{VV}}\Ket{V_{s}V_{i}}\Bra{H_{s}H_{i}})\nonumber
\end{align}
with associated tangle and linear entropy, which are often used to determine the quality of entanglement sources \cite{coffman2000distributed,PhysRevA.61.040101}, given by $T=|\Braket{\phi_{HH}|\phi_{VV}}|^{2}$ and $L=\frac{2}{3}(1-|\Braket{\phi_{HH}|\phi_{VV}}|^{2})$, respectively. For sources pumped by femtosecond lasers, small temporal differences in the two paths is a particular issue \cite{PhysRevA.62.011802} as the photons' coherence time is short. One way to overcome this problem is to filter the photons through a narrow bandpass filter, thus imposing temporal and spectral matching between the $H$ and $V$ photons that pass through. However, such a solution also degrades both the efficiency and reliability of the source. Instead, we demonstrate the ability to compensate for temporal differences without losing any photons.

\section{Experiment}

\begin{figure}[htbp]
\centering\includegraphics[width=8.4cm,keepaspectratio]{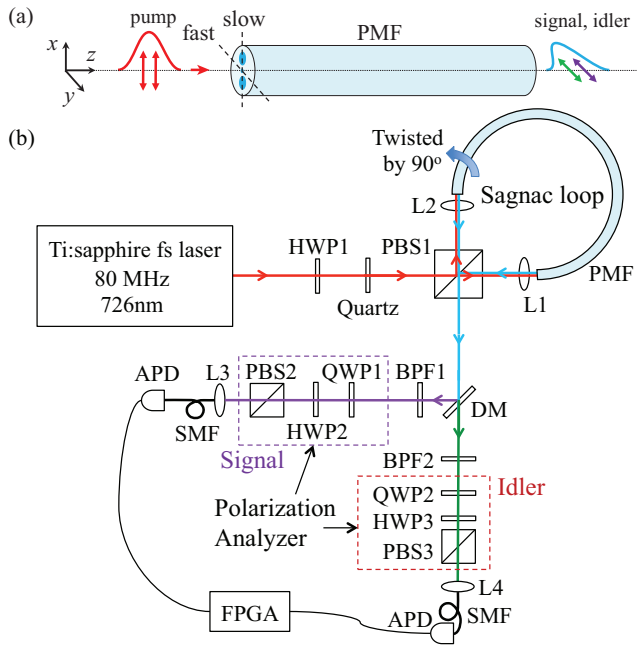}\caption{(a) Schematic of photon-pair generation in standard polarization-maintaining fiber. (b) Experimental setup. M, mirror; L1-4 lens; HWP1-3, half-wave-plate; QWP1-2, quarter-wave-plate; PBS1-3, polarizing beam splitter; PMF, polarization-maintaining fiber; DM, dichroic mirror; BPF1-2, band-pass filter; SMF, single-mode fiber; APD, avalanche photodiode; FPGA, field programmable gate array.\label{exp}}
\end{figure}

The experimental setup is shown in Fig.~\ref{exp}(b). The pump beam originates from a $80\,\text{MHz}$ Ti:sapphire femtosecond laser working at $726\,\text{nm}$ central wavelength that generates pulses with approximately $6\,\text{nm}$ full-width at half-maximum (FWHM) bandwidth. A half-wave-plate (HWP1) sets the polarization of the pump at $45^{\circ}$. The horizontal and vertical polarization portions of the pump are balanced to a power of $\sim 5\,\text{mW}$ each and coupled into opposite ends of a $20\,\text{cm}$-long PMF (Nufern PM630-HP), which is angle-polished to reduce back-reflections. Inside the Sagnac loop, two sideband photons are created, the signal at $634\,\text{nm}$ and the idler at $850\,\text{nm}$. The entangled state is generated at the output port of PBS1. In this configuration the pump is partially suppressed, as it is reflected back through the input port. A dichroic mirror (DM) (Semrock FF685-Di02) reflects the signal photons and transmits the idler photons. A bandpass filter (BPF1) (Semrock FF01-630/20) at the signal arm is used to suppress the pump and other background while another bandpass filter (BPF2) (Semrock FF01-832/37) does the same at the idler arm. Each arm then goes through a polarization analyzer composed of a quarter-wave-plate (QWP), a half-wave-plate (HWP) and a PBS. The photons transmitted through PBS2 and PBS3 are coupled into SMFs connected to silicon-based avalanche photodiodes (APDs) that detect the photons (Excelitas SPCM AQ4C). The electronic output of the APDs is analyzed by a coincidence counter based on a field-programmable gate array (FPGA) \cite{:FPGA} connected to a personal computer.

\section{Results and discussion}

To fully characterize the entangled state, we perform a quantum state tomography \cite{james2001measurement} through 36 different settings of HWP2, HWP3, QWP1 and QWP2 that project the photons onto different joint combinations of the $\{H,V\}$; $\{D=(H+V)/\sqrt{2},A=(H-V)/\sqrt{2}\}$; and $\{L=(H-iV)/\sqrt{2},R=(H+iV)/\sqrt{2}\}$ bases and obtain coincidences statistics. For each setting of the wave-plates, we count coincidences for 15 seconds. Using the maximum likelihood method, we reconstruct the density matrix of the state and determine the tangle and linear entropy. 

\begin{figure}[htbp]
\centering\includegraphics[width=8.4cm,keepaspectratio]{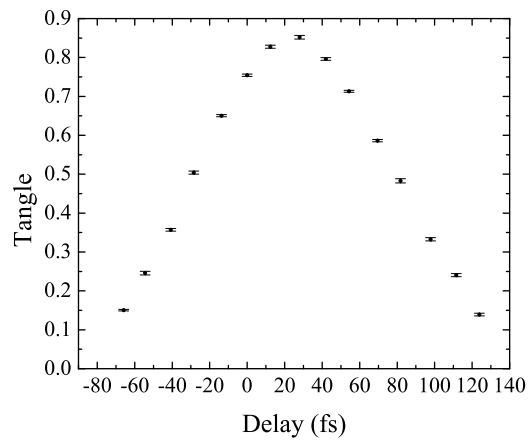}
\caption{Tangle (black dots) versus delay imposed on the vertical polarization of the pump relative to the horizontal polarization. Error bars are estimated assuming Poissonian statistics of the counts.\label{tangle}}
\end{figure}

\begin{figure}[htbp]
\centering\includegraphics[width=8.4cm,keepaspectratio]{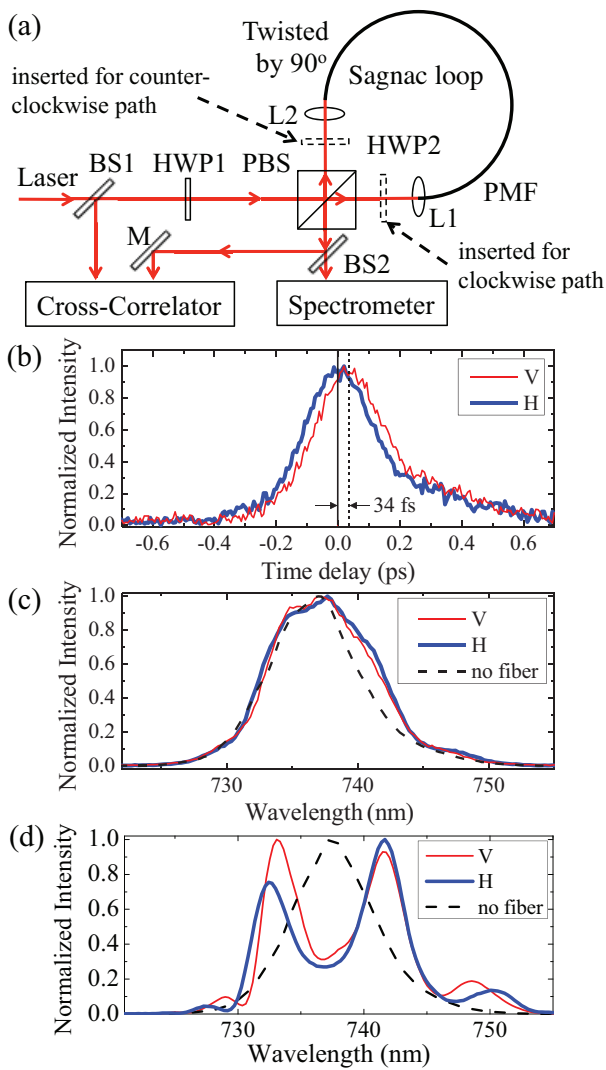}
\caption{(a) Schematic of the measurement of intensity cross-correlations of a reference portion of the pump before entering the fiber with the pump emerging out of the clockwise and counter-clockwise paths of the interferometer. M, mirror; BS1-2, beamsplitter; L1-2 lens; HWP1-2, half-wave-plate; PBS, polarizing beam splitter; PMF, polarization-maintaining fiber. The power coupled into each path is $\sim 5\,\text{mW}$. (b) Cross-correlations of the reference with the clockwise (red, $V$ polarization) and counter-clockwise (blue, H polarization) paths. The cross-correlations indicate the clockwise path is $\sim 34\,\text{fs}$ shorter than the counter-clockwise path. (c) Spectra of the pumps emerging from the fiber for $5\,\text{mW}$ pump power. The solid red ($V$) and blue ($H$) lines indicate the clockwise and counter-clockwise paths, respectively. Comparing to the pump spectrum without the fiber in the Sagnac loop (dashed black), we can see that spectral reshaping has occurred. (d) Spectra of the pumps emerging from the fiber for $50\,\text{mW}$ pump power.  The spectral reshaping becomes more pronounced and the two paths differ from each other to a greater extent at this higher pump power.\label{crosscorr}}
\end{figure}

As noted above, the quality of the entanglement or linear entropy can be reduced due to distinguishability between the two paths in which the photons are created. Since PMF supports only a single spatial mode, the photons are created in that mode, and thus the spatial degree of freedom is uncorrelated with the spectral-temporal modes, while the coupling into SMFs ensures the spatial overlap of photons generated in the counter-propagating paths. We investigate the effect of temporal distinguishability on the quality of the entangled-photons source: using various combinations of three quartz crystals that produce 13, 41, and $68\,\text{fs}$ delays between orthogonal polarizations, we varied the delay between the orthogonally polarized pumps and determined the tangle at each delay, shown in Fig.~\ref{tangle}. We achieve the highest tangle $T_{\text{max}}=0.8522\pm0.0047$ (all errors are estimated assuming Poissonian statistics of the counts) when the vertically polarized pump is delayed by $28\,\text{fs}$ with respect to the horizontally-polarized pump. The associated linear entropy for this case is $L_\text{min}=0.1021\pm0.0030$. For comparison, when no delay is imposed, we find tangle and linear entropy $T=0.7543\pm0.0029$ and $L=0.1649\pm0.0019$, respectively.

In order to understand the sources of distinguishing information between the $H$ and $V$ created photons, we profiled the temporal and spectral shape of the pump pulses emerging from the Sagnac interferometer. Any distinguishing information imposed on the pumps is conveyed to the photon-pairs during the creation process. First, we performed an intensity cross-correlation of a reference portion of the pump taken before entering the interferometer with the clockwise- or counter-clockwise-traveling pump emerging out of the interferometer (see Fig.~\ref{crosscorr}(a)). The cross-correlation measurements indicate the existence of a delay between the pump pulses emerging from different paths. We find that this delay is a few tens of femtoseconds and can vary between measurements depending on slight changes in setup alignment. An example of a cross-correlation measurement is shown in Fig.~\ref{crosscorr}(b). These findings emphasize the need for careful attention when using femtosecond pulses, even in a setup that is nominally perfectly symmetric, such as the Sagnac interferometer. The symmetry of the system may be broken due to imperfect spatial overlap between the incoming pump beam -- which is set by alignment -- and the mode coupled into the PMF. 

We further looked into distinguishing information that may arise due to asymmetric self-phase-modulation effects; we compared the spectra of the emerging pumps (Fig.~\ref{crosscorr}(c)), and found that spectral reshaping occurs even down to low input powers of $\sim 5\,\text{mW}$, albeit the spectra of the two counter-propagating pumps is similar. We therefore conclude that even at the low powers used in this experiment, the self-phase-modulation imposed by the pump on itself results in spectral reshaping of the pulses. The fact that the reshaping is symmetric (\textit{i.e.}, similar for both paths) implies that spectral reshaping does not play a major role in introducing distinguishing information between the two paths for the pump power used in the experiment. However, for higher pump powers, we find that the self-phase-modulation results in path-dependent spectral reshaping (see Fig.~\ref{crosscorr}(d)), which may introduce distinguishing information between the $H$ and $V$ created photon pairs and thus reduce the tangle of the generated state.

\begin{figure}[htbp]
\centering\includegraphics[width=8.4cm,keepaspectratio]{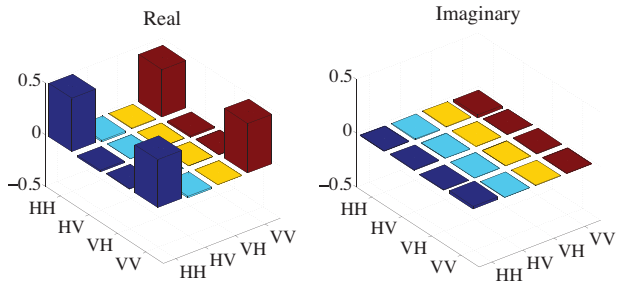}
\caption{Real and imaginary parts of the reconstructed density matrix with $28\,\text{fs}$ temporal compensation. Without any corrections or background subtractions the fidelity to the Bell state $\Ket{\Phi^{+}}=(\Ket{H_{s}H_{i}}+\Ket{V_{s}V_{i}})/\sqrt{2}$ is calculated as $95.86\pm0.10\%$.\label{density_matrix}}
\end{figure}

\begin{figure}[htbp]
\centering\includegraphics[width=8.4cm,keepaspectratio]{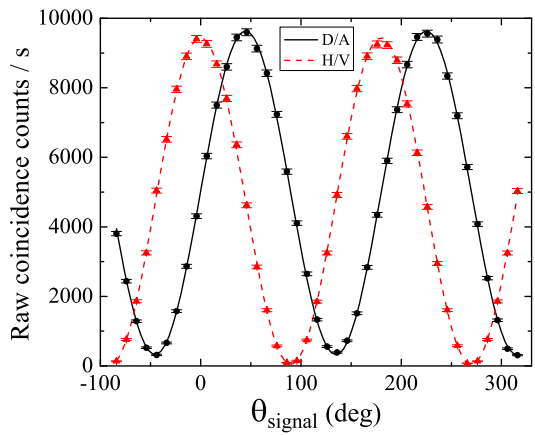}
\caption{Quantum interference visibility in the H/V (red triangles) and D/A (black circles) bases. $\theta_{\mathrm{signal}}$ is the angle by which  the signal arm polarization is rotated.\label{visibility}}
\end{figure}

For the case of maximal tangle, we tilted one of the quartz crystals to control and set the phase $\varphi=0$. The reconstructed density matrix (Fig.~\ref{density_matrix}) suggests $95.86\pm0.10\%$ fidelity with the Bell state $(\Ket{H_{s}H_{i}}+\Ket{V_{s}V_{i}})/\sqrt{2}$. We validate the generated state by an independent quantum-interference visibility measurement: QWP1 and QWP2 are taken out for this measurement. HWP3 is fixed at either $0^{\circ}$ for the $H/V$ basis or $22.5^{\circ}$ for the $D/A$ basis interference measurement. HWP2 rotates the signal polarization by an angle $\theta$, and the coincidences between signal and idler are recorded. The results (see Fig.~\ref{visibility}) show a visibility of $99.22\pm0.46\%$ in the $H/V$ basis and $93.43\pm0.60\%$ in the $D/A$ basis (again, without any corrections or background subtractions). This is consistent with the expected visibilities given the density matrix in Fig.~\ref{density_matrix}, which are $99.12\%$ and $92.92\%$, respectively.

\section{Conclusion}

In summary, we have demonstrated a new polarization-entangled photon-pair source using a standard polarization-maintaining fiber with birefringent phase matching. By exploring the influence of temporal distinguishability, we were able to obtain entanglement comparable in quality to other demonstrated sources while avoiding spectral filtering. The produced photons are practically free of Raman contamination even at room temperature operation owing to the large spectral detuning from the pump, which is enabled by the large birefringence of the fiber. The ability to engineer the source such that the entangled photons are spectrally uncorrelated \cite{PhysRevA.83.031806,Fang:OptExpress:2013} gives the possibility to create photon pairs entangled only in the polarization domain, a crucial requirement for many quantum applications such as teleportation, entanglement swapping and linear optical quantum computing. The source's spatial mode makes it promising for incorporation in free space or fiber networks, while the high tunability \cite{Smith:OptExpress:2009} over the visible wavelengths allows matching to demonstrated quantum memories. We expect this source to be a good candidate for future implementations of quantum information protocols.

\begin{acknowledgments}
This work was supported in part by the University of Delaware Research Foundation and the NSF Physics Division, Grant No.~1205812.
\end{acknowledgments}



\begin{thebibliography}{99}
\newcommand{\enquote}[1]{``#1''}

\bibitem{PhysRevLett.67.661}
A.~K. Ekert, \enquote{Quantum cryptography based on {B}ell's theorem,} Phys.
  Rev. Lett. \textbf{67}, 661--663 (1991).

\bibitem{knill2001scheme}
E.~Knill, R.~Laflamme, and G.~J. Milburn, \enquote{A scheme for efficient
  quantum computation with linear optics,} Nature \textbf{409}, 46--52 (2001).

\bibitem{weihs2001photonic}
W.~Tittel and G.~Weihs, \enquote{Photonic entanglement for fundamental tests
  and quantum communication,} Quantum Inf. Comput. \textbf{1}, 3--56 (2001).

\bibitem{PhysRevLett.75.4337}
P.~G. Kwiat, K.~Mattle, H.~Weinfurter, A.~Zeilinger, A.~V. Sergienko, and
  Y.~Shih, \enquote{New high-intensity source of polarization-entangled photon
  pairs,} Phys. Rev. Lett. \textbf{75}, 4337--4341 (1995).

\bibitem{PhysRevA.74.013802}
J.~F. Hodelin, G.~Khoury, and D.~Bouwmeester, \enquote{Optimal generation of
  pulsed entangled photon pairs,} Phys. Rev. A \textbf{74}, 013802 (2006).

\bibitem{kuzucu2008pulsed}
O.~Kuzucu and F.~N.~C. Wong, \enquote{Pulsed sagnac source of narrow-band
  polarization-entangled photons,} Phys. Rev. A \textbf{77}, 032314 (2008).

\bibitem{sauge2007narrowband}
S.~Sauge, M.~Swillo, S.~Albert-Seifried, G.~B. Xavier, J.~Waldeb\"{a}ck,
  M.~Tengner, D.~Ljunggren, and A.~Karlsson, \enquote{Narrowband
  polarization-entangled photon pairs distributed over a wdm link for qubit
  networks,} Opt. Express \textbf{15}, 6926--6933 (2007).

\bibitem{Hubel:OptExpress:2007}
H.~H\"{u}bel, M.~R. Vanner, T.~Lederer, B.~Blauensteiner, T.~Lor\"{u}nser,
  A.~Poppe, and A.~Zeilinger, \enquote{High-fidelity transmission of
  polarization encoded qubits from an entangled source over 100 km of fiber,}
  Opt. Express \textbf{15}, 7853--62 (2007).

\bibitem{takesue2005generation}
H.~Takesue, K.~Inoue, O.~Tadanaga, Y.~Nishida, and M.~Asobe,
  \enquote{Generation of pulsed polarization-entangled photon pairs in a
  1.55-$\mu$m band with a periodically poled lithium niobate waveguide and an
  orthogonal polarization delay circuit,} Opt. Lett. \textbf{30}, 293--295
  (2005).

\bibitem{lim2008stable}
H.~C. Lim, A.~Yoshizawa, H.~Tsuchida, and K.~Kikuchi, \enquote{Stable source of
  high quality telecom-band polarization-entangled photon-pairs based on a
  single, pulse-pumped, short {PPLN} waveguide,} Opt. Express \textbf{16},
  12460--12468 (2008).

\bibitem{zhong2010high}
T.~Zhong, X.~Hu, F.~N.~C. Wong, K.~K. Berggren, T.~D. Roberts, and P.~Battle,
  \enquote{High-quality fiber-optic polarization entanglement distribution at
  1.3 $\mu$m telecom wavelength,} Opt. Lett. \textbf{35}, 1392--1394 (2010).

\bibitem{Martin:NewJournalOfPhysics:2010}
A.~Martin, A.~Issautier, H.~Herrmann, W.~Sohler, D.~B. .~B. Ostrowsky,
  O.~Alibart, and S.~Tanzilli, \enquote{A polarization entangled photon-pair
  source based on a type-ii ppln waveguide emitting at a telecom wavelength,}
  New J. Phys. \textbf{12}, 103005 (2010).

\bibitem{Kaiser:NewJournalOfPhysics:2012}
F.~Kaiser, A.~Issautier, L.~A. .~A. Ngah, O.~D\u{a}nil\u{a}, H.~Herrmann,
  W.~Sohler, A.~Martin, and S.~Tanzilli, \enquote{High-quality polarization
  entanglement state preparation and manipulation in standard telecommunication
  channels,} New J. Phys. \textbf{14}, 085015 (2012).
  
\bibitem{PhysRevA.76.043836}
J.~Fan, M.~D. Eisaman, and A.~Migdall, \enquote{Bright Phase-stable Broadband Fiber-based Source of Polarization-entangled Photon Pairs.} 
  Phys. Rev. A \textbf{76}, 043836 (2007).


\bibitem{PhysRevLett.99.120501}
J.~Fulconis, O.~Alibart, J.~L. O'Brien, W.~J. Wadsworth, and J.~G. Rarity,
  \enquote{Nonclassical interference and entanglement generation using a
  photonic crystal fiber pair photon source,} Phys. Rev. Lett. \textbf{99},
  120501 (2007).

\bibitem{ling2009mode}
A.~Ling, J.~Chen, J.~Fan, and A.~Migdall, \enquote{Mode expansion and {B}ragg
  filtering for a high-fidelity fiber-based photon-pair source,} Opt. Express
  \textbf{17}, 21302--21312 (2009).

\bibitem{PhysRevA.70.031802}
H.~Takesue and K.~Inoue, \enquote{Generation of polarization-entangled photon
  pairs and violation of bell's inequality using spontaneous four-wave mixing
  in a fiber loop,} Phys. Rev. A \textbf{70}, 031802 (2004).
  
\bibitem{Li:PhysicalReviewLetters:2005}
X.~Li, P.~Voss, J.~Sharping, and P.~Kumar, \enquote{Optical-fiber source of
  polarization-entangled photons in the 1550nm telecom band,} Phys. Rev. Lett.
  \textbf{94}, 053601 (2005).

\bibitem{Medic:OptLett:2010}
M.~Medic, J.~B. Altepeter, M.~A. Hall, M.~Patel, and P.~Kumar,
  \enquote{Fiber-based telecommunication-band source of degenerate entangled
  photons,} Opt. Lett. \textbf{35}, 802--4 (2010).

\bibitem{zhoupolarization}
Q.~Zhou, W.~Zhang, P.~Wang, Y.~Huang, and J.~Peng, \enquote{Polarization
  entanglement generation at 1.5 $\mu$m based on walk-off effect due to fiber
  birefringence,} Opt. Lett \textbf{37}, 1679--1681 (2012).
  
 \bibitem{Zhou:Arxiv13077207QuantPh:2013}
Q.~Zhou, W.~Zhang, T. Niu, S.~Dong, Y. Huang, and J. Peng, \enquote{A
  polarization maintaining scheme for 1.5 mmmmm polarization entangled photon
  pair generation in optical fibers,} arXiv:1307.7207 [quant-ph]  (2013).

  
\bibitem{hall2009drop}
M.~A. Hall, J.~B. Altepeter, and P.~Kumar, \enquote{Drop-in compatible
  entanglement for optical-fiber networks,} Opt. Express \textbf{17},
  14558--14566 (2009).

\bibitem{Hall:PhysicalReviewLetters:2011}
M.~A. Hall, J.~B. Altepeter, and P.~Kumar, \enquote{Ultrafast switching of
  photonic entanglement,} Phys. Rev. Lett. \textbf{106}, 053901 (2011).

\bibitem{Smith:OptExpress:2009}
B.~J. Smith, P.~Mahou, O.~Cohen, J.~S. Lundeen, and I.~A. Walmsley,
  \enquote{Photon pair generation in birefringent optical fibers,} Opt. Express
  \textbf{17}, 23589--602 (2009).

\bibitem{stolen1981phase}
R.~H. Stolen, M.~A. Bosch, and C.~Lin, \enquote{Phase matching in birefringent
  fibers,} Opt. Lett. \textbf{6}, 213--215 (1981).
  
\bibitem{Fang:OptExpress:2013}
B.~Fang, O.~Cohen, J.~B.~Moreno, and V.~O.~Lorenz, \enquote{State engineering of photon pairs produced through dual-pump spontaneous four-wave mixing,} Opt. Express \textbf{21}, 2707--2717 (2013).

\bibitem{PhysRevA.83.031806}
C.~S\"{o}ller, O.~Cohen, B.~J. Smith, I.~A. Walmsley, and C.~Silberhorn,
  \enquote{High-performance single-photon generation with commercial-grade
  optical fiber,} Phys. Rev. A \textbf{83}, 031806 (2011).

\bibitem{Simon2010}
C.~Simon, M.~Afzelius, J.~Appel, A.~Boyer de~la Giroday, S.~J. Dewhurst,
  N.~Gisin, C.~Y. Hu, F.~Jelezko, S.~Kr�ll, J.~H. M�ller, J.~Nunn, E.~S.
  Polzik, J.~G. Rarity, H.~De~Riedmatten, W.~Rosenfeld, A.~J. Shields,
  N.~Sk�ld, R.~M. Stevenson, R.~Thew, I.~A. Walmsley, M.~C. Weber,
  H.~Weinfurter, J.~Wrachtrup, and R.~J. Young, \enquote{Quantum memories,}
  Eur. Phys. J. D \textbf{58}, 1--22 (2010).

\bibitem{Hammerer2010}
K.~Hammerer, A.~S. S\o{}rensen, and E.~S. Polzik, \enquote{Quantum interface
  between light and atomic ensembles,} Rev. Mod. Phys. \textbf{82}, 1041--1093
  (2010).
  
 \bibitem{meyer2012generating}
E.~Meyer-Scott, V.~Roy, J.-P. .~P. Bourgoin, B.~L. Higgins, L.~K. Shalm, and
  T.~Jennewein, \enquote{Generating polarization-entangled photon pairs using
  cross-spliced birefringent fibers,} Opt. Express \textbf{21}, 6205--6212
  (2013).

\bibitem{PhysRevA.69.041801}
M.~Fiorentino, G.~Messin, C.~E. Kuklewicz, F.~N.~C. Wong, and J.~H. Shapiro,
  \enquote{Generation of ultrabright tunable polarization entanglement without
  spatial, spectral, or temporal constraints,} Phys. Rev. A \textbf{69}, 041801
  (2004).

\bibitem{harvey2003scalar}
J.~D. Harvey, R.~Leonhardt, S.~Coen, G.~K.~L. Wong, J.~C. Knight, W.~J.
  Wadsworth, and P.~St~J~Russell, \enquote{Scalar modulation instability in the
  normal dispersion regime by use of a photonic crystal fiber,} Opt. Lett.
  \textbf{28}, 2225--2227 (2003).

\bibitem{coffman2000distributed}
V.~Coffman, J.~Kundu, and W.~K. Wootters, \enquote{Distributed entanglement,}
  Phys. Rev. A \textbf{61}, 052306 (2000).

\bibitem{PhysRevA.61.040101}
S.~Bose and V.~Vedral, \enquote{Mixedness and teleportation,} Phys. Rev. A
  \textbf{61}, 040101 (2000).

\bibitem{PhysRevA.62.011802}
Y.-H. Kim, S.~P. Kulik, and Y.~Shih, \enquote{High-intensity pulsed source of
  space-time and polarization double-entangled photon pairs,} Phys. Rev. A
  \textbf{62}, 011802 (2000).

\bibitem{:FPGA}
\enquote{Fabricated based on a design by the group of {A}. {S}teinberg,
  {U}niversity of {T}oronto,}
  \url{http://www.physics.utoronto.ca/~astummer/pub/mirror/Projects/Archives/2008\%20Coincidence\%20Counter/Coincidence\%20Counter.html}.

\bibitem{james2001measurement}
D.~F.~V. James, P.~G. Kwiat, W.~J. Munro, and A.~G. White, \enquote{Measurement
  of qubits,} Phys. Rev. A \textbf{64}, 052312 (2001).



\end{thebibliography}

\end{document}